\documentclass[aps,rmp,preprint,groupedaddress,floatfix]{revtex4}
\usepackage{natbib}
\usepackage{genetics}
\usepackage{graphicx}
\usepackage{dcolumn}

\begin{document}

\newcommand{\mon}{\begin{displaymath}}
\newcommand{\moff}{\end{displaymath}}
\newcommand{\sumi}[1]{\sum_{{#1}=-\infty}^{\infty}}
\renewcommand{\b}[1]{\mbox{\boldmath ${#1}$}}
\newcommand{\sumy}{\sum_{\b{y}}}
\newcommand{\sumz}{\sum_{\b{z}}}
\newcommand{\pd}[2]{\frac{\partial {#1}}{\partial {#2}}}
\newcommand{\od}[2]{\frac{d {#1}}{d {#2}}}
\newcommand{\inti}{\int_{-\infty}^{\infty}}
\newcommand{\eon}{\begin{equation}}
\newcommand{\eoff}{\end{equation}}
\newcommand{\e}[1]{\times 10^{#1}}
\newcommand{\chem}[2]{{}^{#2} \mathrm{#1}}
\newcommand{\s}{s}
\newcommand{\zetaexp}{\left( \zeta e^{q \s t} \right)}
\newcommand{\taunuc}{\tau_{nuc}}
\newcommand{\eq}[1]{Eq. (\ref{#1})}\
\newcommand{\eqref}{\eq}
\newcommand{\ev}[1]{\langle #1 \rangle}
\newcommand{\expectation}{\ev}
\newcommand{\mat}[1]{\bf{\mathcal{#1}}}
\newcommand{\fig}[1]{Fig. \ref{#1}}
\renewcommand{\u}{U_b}
\newcommand{\sig}{\sigma}
\renewcommand{\r}{R}
\newcommand{\rms}{SD(s)}
\newcommand{\sci}{s_{ci}}
\newcommand{\rmsci}{SD(s_{ci})}

\title{Clonal Interference, Multiple Mutations, and Adaptation in Large Asexual Populations}

\author{Craig A. Fogle${}^*$}
\author{James L. Nagle${}^*$}
\author{Michael M. Desai${}^{\dagger}$}
\affiliation{${}^*$Department of Physics, University of Colorado at
Boulder, Boulder CO 80305 \\${}^\dagger$Lewis-Sigler Institute for
Integrative Genomics, Princeton University, Princeton NJ 08544 }

\date{April 6, 2008}

\linespread{1.0}

\begin{abstract}
Two important problems affect the ability of asexual populations to
accumulate beneficial mutations, and hence to adapt.  First, clonal
interference causes some beneficial mutations to be outcompeted by
more-fit mutations which occur in the same genetic background.  Second,
multiple mutations occur in some individuals, so even mutations of large
effect can be outcompeted unless they occur in a good genetic background
which contains other beneficial mutations. In this paper, we use a Monte
Carlo simulation to study how these two factors influence the adaptation
of asexual populations.  We find that the results depend qualitatively on
the shape of the distribution of the effects of possible beneficial
mutations. When this distribution falls off slower than exponentially,
clonal interference alone reasonably describes which mutations dominate
the adaptation, although it gives a misleading picture of the
evolutionary dynamics. When the distribution falls off faster than
exponentially, an analysis based on multiple mutations is more
appropriate.  Using our simulations, we are able to explore the limits of
validity of both of these approaches, and we explore the complex dynamics
in the regimes where neither are fully applicable.
\end{abstract}

\maketitle

\linespread{1.6}

\section{Introduction}

The accumulation of beneficial mutations drives adaptation and
evolutionary innovation.  Yet despite its central importance, the
evolutionary dynamics by which a population accumulates such mutations is
poorly understood.  To better understand adaptation in any particular
system, we must ask two questions.  First, what is the range of
beneficial mutations that are possible given the particular environmental
challenge and genetic state of the population? Second, given this set of
possibilities, what will actually happen probabilistically?

The first of these questions is fundamentally empirical, though Orr and
Gillespie have argued on general theoretical grounds that the
distribution of fitness effects of beneficial mutations should be
exponential \cite{orr02, orr03, gillespie83, gillespie84, gillespie91}. A
variety of recent experimental studies are roughly consistent with this
exponential expectation \cite{imhof01, rozen02, kassen06, rokyta05,
sanjuan04, depristo05, lunzer05}. However, beneficial mutations are rare
and their fitness effects are difficult to measure precisely, so these
experimental studies are generally based on relatively few total
mutations and have correspondingly limited resolution. The tail of the
distribution, which refers to the rare mutations which confer a very
large fitness benefit, is particularly hard to measure. Further, the
spectrum of beneficial mutations available to a population is likely to
vary with genetic background, history, and the environment, so it is
unclear how far we can generalize from individual experimental studies.
Thus it is still unknown whether in general the distribution of mutant
effects, particularly of large-effect mutations, is exponential.

Even if we knew the precise distribution of mutational possibilities, it
is not clear how a population would evolve.  Because mutations are random
events, there will inevitably be some randomness in how a given
population adapts.  What we would like to understand is the statistics of
which beneficial mutations are more or less likely to contribute to
adaptation, and the dynamics by which they do so.  That is, given a set
of things that are possible, what is the probability that any given one
of them will actually occur and contribute to the adaptation of the
population? What is the evolutionary dynamics by which they do so? In
this paper, we focus on how the distribution of mutations that actually
occur and spread through the entire population (i.e. \emph{fix}),
$\rho_f(s)$, depends on the distribution of mutations that are possible,
$\rho(s)$, where $s$ is the fitness benefit from a single mutation.  We
explore these features as a function of the population size $N$ and the
overall mutation rate $U$.  Besides its importance in understanding
adaptation, this question is relevant in practical attempts to measure
the distribution of possible mutations, since $\rho_f(s)$ is much easier
to measure experimentally than $\rho(s)$.  We also examine some aspects
of the dynamics by which the mutations that fix do so.

There are a number of effects that make the distribution of mutations
that fix different from the distribution of all possible mutations.
First, most beneficial mutations that occur are lost rapidly by random
genetic drift. If a beneficial mutation is particularly lucky, it will
avoid this stochastic loss and reach a high enough frequency that
thereafter its dynamics become dominated by selection rather then drift.
We refer to this process as the \emph{establishment} of the beneficial
mutation. Mutations of larger effect are more likely to survive random
drift --- they have a higher establishment probability --- so this will
tend to bias the distribution of mutations that actually fix towards
larger-effect mutations, relative to the distribution of mutations that
are possible \cite{haldane27, rozen02}.

Once a mutation has become established, it will fix provided that nothing
else interferes.  However, this fixation takes time, and other beneficial
mutations can become established in individuals without the original
mutation before the original mutation can fix.  In an asexual population,
if one or more of these other mutations has a larger fitness benefit than
the original mutation, the original mutation will eventually be
out-competed and driven to extinction.  This process is known as
\emph{clonal interference} \cite{gerrishlenski98, gerrish01, wilke04}.
The same process also operates in a sexual population, where it is
referred to as the Hill-Robertson effect, but is mitigated because the
two competing mutations can potentially recombine onto the same genome
and fix together \cite{hillrobertson66}. In this paper, we focus
exclusively on asexual populations, where this effect is strongest.

In a small population with a small to modest mutation rate, the
establishment of a beneficial mutation is an extremely rare event. Thus
clonal interference is unlikely to occur, and the distribution of
mutations that fix is simply the distribution of mutations that
establish.  In a larger population, or one with a higher mutation rate,
however, clonal interference can be extremely common. Because
small-effect mutations are more likely to be interfered with than
large-effect mutations, clonal interference biases the distribution of
mutations that fix towards those of large effect. This bias has been
analyzed in detail both theoretically \cite{gerrishlenski98,wilke04} and
experimentally \cite{devisser99, devisserrozen05}.

These analyses of clonal interference only consider mutations which occur
in the wild-type population; they assume that the largest such mutation
is the one that fixes.  The possibility of double mutations in a single
organism is neglected.  But, in fact, even if a more-fit mutation B
occurs before an earlier but less-fit mutation A fixes, A may still
survive, because an individual with mutation A can get another mutation C
such that the A-C double mutant is more fit than B.  Recently,
\citet{desaifisher07} showed that whenever clonal interference is
important, these multiple mutations are also at least of comparable
importance --- and, in fact, many large asexual populations will often
routinely have triple or quadruple mutations \cite{desaimurray07}.
Because small-effect mutations are more common than mutations of larger
effect, they are more likely to form double mutants.  Thus the
possibility of multiple mutations biases the distribution of mutations
that fix back towards those of smaller effect. In short, it will often be
the case that getting two small-effect mutations is more common than
getting a single (rarer) large-effect mutation.  This effect depends on
the shape of the distribution of mutational effects: the rarer
large-effect mutations are compared to small-effect ones, the stronger
the multiple-mutation effect should be. The importance of this effect
also depends on population size and mutation rate, though in a somewhat
different way than clonal interference does.

In addition to affecting the distribution of mutations that fix, multiple
mutations also have an important impact on the evolutionary dynamics.
Different individuals have different numbers and strengths of beneficial
mutations, so a large population can maintain substantial variation in
fitness.  It is only those mutations that occur in the most-fit
individuals which have the best chance of surviving and contributing to
the long-term adaptation of the population.  Thus the dynamics of
adaptation are slowed down, limited by the rate at which good mutations
occur in good backgrounds.

Because the distribution of beneficial mutations which fix depends in a
subtle way on both clonal interference and multiple-mutation effects, it
cannot be fully understood without a complete model which includes both.
No analytical results from such a model yet exist, though
\citet{kimorr05} have analyzed a model which includes some aspects of
both effects. In this paper, we address this question using Monte Carlo
simulations of the full evolutionary dynamics of large asexual
populations, including both clonal interference and multiple-mutation
effects.  We consider several distributions of possible mutational
effects $\rho(s)$, and determine the distribution of mutations that fix,
$\rho_f(s)$, across a range of population sizes and mutation rates. We
find that clonal interference analysis provides a good approximation for
some aspects of $\rho_f(s)$ when large-effect mutations are sufficiently
common relative to small-effect ones.  When large-effect mutations are
more rare, we find that an approximation focusing on multiple mutation
effects, proposed by \citet{desaifisher07}, can be more appropriate.

We next turn to the evolutionary dynamics by which beneficial mutations
fix. Using our Monte Carlo approach, we simulate the dynamics of the full
model with a distribution of fitness effects. We show that
multiple-mutation dynamics involving mutations within a narrow range of
fitness effects describes the evolution.  We describe how this range of
fitness effects depends on $\rho(s)$ and the other parameters, and how
the mutations within this range accumulate.

\section{Model and Simulation Methods}

We consider an asexual population of $N$ haploid individuals with an
overall mutation rate $\u$ towards beneficial mutations.  Our model also
applies to asexual diploids, where the fitness effects of mutations refer
to their effects in the individual in which they occur. Given that a
beneficial mutation occurs, we assume that its fitness effect is $s$
(i.e. the fitness of the organism increases by a factor of $e^{s}$) with
probability \eon \rho(s) = \frac{e^{-(s/\sig)^\beta}}{\sig \Gamma (1 +
1/\beta)}, \eoff where the two parameters $\sig$ and $\beta$ characterize
the shape of the distribution and $\Gamma$ is the Gamma function.  This
form for $\rho(s)$ allows us to explore the importance of the shape of
the tail of the distribution of mutant effects --- that is, the relative
rareness of large-effect mutations compared to small-effect ones.  When
$\beta = 1$, the distribution of mutant effects is exponential with mean
$\sig$. When $\beta > 1$, the distribution of mutant effects falls off
faster than exponentially (i.e. large-effect mutations are more rare),
and the effect of a ``typical'' beneficial mutation is $\sig$. When
$\beta < 1$ the distribution of mutation effects falls off more slowly
than exponentially (large-effect mutations are more common), and again
the effect of a typical beneficial mutation is $\sig$. Note that this
distribution is normalized to $1$, so beneficial mutations with effect
between $s$ and $s+ds$ occur at a rate $\u \rho(s) ds$.  The average
fitness effect of a beneficial mutation, $\ev{s}$, is \eon \ev{s} = \sig
\frac{\Gamma (2/\beta)}{\Gamma (1/\beta)}, \eoff so while $\sig$ is
always a typical effect of a beneficial mutation, it is only exactly
equal to the average effect for $\beta = 1$. We assume that there is no
epistasis, so that an individual with two mutations of effect $s_1$ and
$s_2$ has fitness $e^{s_1+s_2}$ (or more generally for n mutations, the
fitness is $\prod_{i=1}^{n} e^{s_{i}}$). We neglect deleterious
mutations, as these are not expected to qualitatively affect the dynamics
in large populations when beneficial mutations are relatively common,
which is the situation we study \cite{rouzine03, desaifisher07}.

We assume a dynamics with discrete generations.  In each generation, we
first randomly select which individuals will survive to the next
generation, weighted by each individual's fitness.  The overall survival
probabilities are normalized so that on average half of the population
will survive to the next generation.  Each surviving individual then
duplicates to create two identical individuals in the next generation.
Finally, each of these individuals in the next generation has a
probability $\u$ of acquiring a new beneficial mutation. We then repeat
this algorithm for the subsequent generation.  We record all the
information about the genetic state of the population at each step. All
simulations were checked to ensure that the results were extracted after
a steady-state had been achieved.  Note that this algorithm does not
enforce an exact population size $N$ at each step, but rather keeps the
average population size equal to $N$.

These evolutionary dynamics have the advantage of being fast to simulate,
allowing us to explore a greater range of parameters in a reasonable
amount of computation time.  However, they are slightly different from
standard Wright-Fisher dynamics.  It is not clear which of these models
is the best representation of any particular population, but our model
might be expected to correspond to a bacterial or yeast population which
divides by binary fission.  The differences between models do lead to
small differences in establishment probabilities, and thus are expected
to cause minor modifications to the evolutionary dynamics. However, we
have also simulated the standard Wright-Fisher dynamics for many of the
parameters we describe in this paper, and the results are all
qualitatively similar.

\section{Results}

\subsection{The Distribution of Mutations that Fix}

We begin by studying the distribution of mutations that fix, $\rho_f(s)$,
as compared to the distribution of mutations that occur, $\rho(s)$. When
$N$ and $\u$ are sufficiently small, neither clonal interference nor
multiple mutations will occur.  Thus the distribution of mutations that
fix equals the distribution of mutations that establish.  We expect \eon
\rho_f(s) = \pi(s) \rho(s), \eoff valid for small $N$ and $\u$, where
$\pi(s)$ is the establishment probability. Note that by convention we
will assume $\rho_f(s)$ is \emph{not} normalized; this will make various
calculations much more transparent. In our model, this establishment
probability is given by \eon \pi (s) = \frac{1-e^{-2s}}{1-e^{-2Ns}},
\eoff assuming that $s$ is the fitness advantage relative to the
background population.  When $Ns \ll 1$ we have $\pi(s) \approx
\frac{1}{N}$, while for $Ns \gg 1$ but $s \ll 1$ we have $\pi(s) \approx
2s$. This small-population limit of $\rho_f(s)$, which is the
distribution of mutations that establish, is sometimes called the
distribution of \emph{contending} mutations \cite{rozen02}.  We will
denote it by $\rho_c(s)$.

In larger populations, we expect clonal interference to suppress the
fixation of small-effect mutations.  Thus for small $s$, we expect
$\rho_f(s)$ to be smaller than $\rho_c(s)$.  But clonal interference
cannot suppress the fixation of the largest-effect mutations, so if it
were the only important effect we would expect that for large $s$,
$\rho_f(s)$ should equal $\rho_c(s)$. Provided that small-effect
mutations are common enough relative to those of larger effect, however,
multiple small-$s$ mutations can suppress the fixation of large-$s$
mutations. \citet{desaifisher07} suggest that this will occur whenever
the distribution of mutation effects falls off faster than exponentially
(i.e. when $\beta > 1$). When this is true, we expect that these multiple
mutation effects will cause $\rho_f(s)$ to be smaller than $\rho_c(s)$
even for large $s$. When $\beta \leq 1$, large-effect mutations will
typically not be out-competed by multiple smaller-effect mutations, so
$\rho_f(s) = \rho_c(s)$ for large $s$, though multiple mutations will
often still be important to the dynamics of adaptation.

In Fig. 1, we show several examples of the distribution of fixed
mutations $\rho_f(s)$ compared to the distribution of possible and
contending mutations, $\rho(s)$ and $\rho_c(s)$ respectively.  The
predictions of clonal interference analysis are also shown (see
Discussion). We see that the above expectations are met. For small
populations (Fig. 1a), $\rho_f(s) = \rho_c(s)$ except for very small $s$
(these $s$ are so small that clonal interference prevents them from
fixing even in these small populations). We find similar results for
small populations regardless of $\beta$ (data not shown). For larger
populations, the behavior depends on $\beta$.  In all cases, small-effect
mutations are suppressed quite dramatically by clonal interference
effects (Fig. 1b-d).  Note, however, that clonal interference analysis
predicts that this suppression of small-effect mutations should be even
stronger than we observe.  Presumably this is because some of these
small-effect mutations are able to fix in multiple-mutation combinations
together with those of larger effect.  For large $\beta$, the
largest-effect mutations are also suppressed, presumably by
multiple-mutation effects (Fig. 1b).

\subsection{Which Mutations Contribute To Adaptation}

To determine how beneficial mutations of different effects contribute to
the overall adaptation of the population, we need to weight mutations by
their fitness effects.  That is, a single fixed mutation with effect $2s$
contributes the same amount to the adaptation of the population as two
mutations with effect $s$.  Thus $\r(s) \equiv s \rho_f(s)$ is the
distribution of the relative contributions to the overall adaptation as a
function of $s$.  The integral of $\r (s)$ from $s_1$ to $s_2$ is the
total contribution of mutations of size between $s_1$ and $s_2$ to the
adaptation.

Using $\r(s)$, we can study which mutations are most important. We expect
that mutations of very small effect will not contribute substantially to
adaptation, because they are strongly suppressed by clonal interference
and do not contribute much even when they do fix.  On the other hand,
mutations of very large effect will be too rare to contribute
substantially, and may be impeded by multiple smaller mutations.  Thus we
expect that typically $\r(s)$ will have a peak at some intermediate value
of $s$, with some range of mutations around this that contribute
substantially to adaptation.  This is indeed what we find. We
characterize $\r(s)$ by its mean, which we call $\tilde s$. In practice,
this mean is a good estimate of the ``peak'' of $\r(s)$. We estimate the
width of the range of mutations around the mean which contribute
substantially to the evolution by the standard deviation of $\r(s)$
around $\tilde s$, which we call $\rms$.

In Fig. 2, we show how $\tilde s$ in our simulated populations depends on
$N$ and $\u$ for several different values of $\beta$.  We compare these
simulated results to the predictions of clonal interference analysis
alone and, for $\beta > 1$, the analysis of \citet{desaifisher07}, which
combines elements of clonal interference and multiple mutations in an
ad-hoc way (see Discussion).

In Fig. 3, we show $\rms / \tilde s$ as a function of $N$ and $\u$, again
for several values of $\beta$, compared to the predictions of clonal
interference analysis. Note that $\rms / \tilde s \ll 1$ corresponds to
the case where a narrow range of mutations around $\tilde s$ dominate the
evolution.  We see that this is typically the case, regardless of whether
the distribution of mutational effects falls off faster or more slowly
than exponentially, except for small $N$ or $\u$.

\subsection{The Dynamics of Adaptation}

We now turn to the dynamics by which these important mutations
accumulate.  We have seen that multiple mutations are important in
determining $\tilde s$ for $\beta > 1$, but not so important for $\beta
\leq 1$.  Despite this, mutations of effect around $\tilde s$, which
dominate the overall adaptation of the population, may accumulate via
multiple-mutations dynamics even when $\beta \leq 1$, although mutations
of much larger effect fix whenever they establish.

These multiple-mutations dynamics are easier to understand intuitively in
an idealized model where all mutations confer the \emph{same} fitness
advantage $s$. If this were the case, we could describe the state of the
population as a distribution of the number of such mutations each
individual has. Some lucky most-fit individuals have more than the
average number of such mutations, and it is only additional mutations
within this small subpopulation that will contribute to the long-term
evolution of the population; others will eventually go extinct because
they are handicapped by their relatively poor genetic background. We
define the \emph{lead}, $q$, to be the number of such mutations the
most-fit individual possesses in excess of the average individual (more
precisely, $q-1$ is defined to be the difference in number of mutations
between the most-fit class of established individuals and the mean
individual).  Several studies have analyzed the accumulation of
beneficial mutations when multiple mutants are important (i.e. $q > 1$),
and found that these multiple mutations dramatically affect the
evolutionary dynamics \cite{desaifisher07,rouzine03,rouzinewilke07}.

When mutations have a range of fitness effects, the dynamics are clearly
more complex. Yet as we have seen, mutations in some narrow range around
some $\tilde s$ dominate the evolution.  It is therefore natural to
expect that multiple mutations of effect of order $\tilde s$ may
routinely appear, and that their accumulation be described roughly by the
single-$s$ model, with $s$ chosen to be $\tilde s$, and the mutation rate
$\tilde \u$ the overall mutation rate to mutations within roughly $\pm$$\rms$
of $\tilde s$. We can define the lead $q$ in this more complex situation
as the fitness of the most-fit individual minus the fitness of the mean
individual, divided by $\tilde s$.  That is, $q$ is the number of extra
$\tilde s$-sized mutations that the most-fit individual has compared to
the average individual (more precisely, we define $q-1$ as the fitness of
the most-fit established class minus the average fitness, divided by
$\tilde s$).

In Fig. 4, we show how this $q$ depends on $N$ and $\u$ for several
different values of $\beta$.  We see that even for $\beta \leq 1$, these
multiple mutations are important to the dynamics (where values of $q=3$
to $5$ are reached). Note that the original clonal interference analysis
implicitly assumes no multiple mutations.  Thus it implies that $q$ is
between $1$ and $2$, depending on whether an established mutant group is
sweeping to fixation.

As beneficial mutations accumulate, the population adapts. We define the
rate of adaptation, $v$, to be the rate at which the average fitness of
the population increases. In Fig. 5, we show how this $v$ depends on
$N$ and $\u$, again for several values of $\beta$. We compare these to
the predictions of clonal interference theory alone and to the analysis
of \citet{desaifisher07} (see Discussion).

\section{Discussion}

Our Monte Carlo simulation approach allows us to study the evolutionary
dynamics of adaptation in large asexual populations, where the effects of
both clonal interference and multiple mutations interact in a subtle way.
Analytic results are difficult to obtain in this complex situation, and
hence such analysis has been confined to approximations that focus
primarily on one or the other effect.  Using our simulations, we can
assess the usefulness and generality of these methods.

In its original form, which neglects the possibility of multiple
mutations, clonal interference analysis predicts the distribution of
mutations that contribute to adaptation, and the dynamics by which they
do so \cite{gerrishlenski98, wilke04}.  In this analysis the probability
that a beneficial mutation fixes is the probability that it establishes
and then fixes before another more-fit mutation establishes.
\citet{gerrishlenski98} found that given that a mutation of effect $s$
has established, the expected number of more-fit mutations establishing
before the original mutation fixes is roughly \eon \lambda(s) \approx
\frac{N \u}{s} \ln N \int_s^{\infty} \pi (x) \rho(x) dx, \label{lambdas}
\eoff assuming that all mutations arise in the wild-type population.
Thus the distribution of mutations that fix is \eon \rho_f(s) = \pi(s)
e^{- \lambda(s)} \rho(s). \label{rhofci} \eoff In Fig. 1, we compare this
prediction to the results of our simulations. From \eq{rhofci} it is also
straightforward to calculate the expected $\tilde s$ and $\rms$. Because
clonal interference becomes more likely as either population size or
mutation rate increases, increasing either of these parameters is
expected to increase $\tilde s$. These predictions are shown in Figs. 2
and 3.

The average rate at which mutations fix equals the rate at which those
destined to fix occur.  Thus clonal interference analysis predicts that
the average fixation rate $\ev{k}$ is \eon \ev{k} = N \u \int_0^\infty
\rho_f(s) ds. \label{clonalfixationrate} \eoff  This means that the rate
of adaptation $v$ is \eon v = \ev{k} \ev{s}, \eoff where $\ev{s}$ is the
average fitness of mutations that fix.  This prediction is shown in Fig.
5.

For $\beta = 0.5$ and $\beta = 1$, we see from Fig. 2 that clonal
interference yields reasonable estimates of $\tilde s$. This makes sense,
as in this regime multiple mutations do not suppress the fixation of
larger-effect mutations. For $\beta = 10$, clonal interference
systematically overestimates $\tilde s$, by an amount which increases
with the population size and mutation rate. This also makes sense, as in
large populations for $\beta = 10$ multiple mutations in fact suppress
the fixation of large-effect mutations.

Although clonal interference accurately predicts $\tilde s$ for small
$\beta$, we can see from Fig. 4 that for all values of $\beta$, $q$ is
greater than $2$ when $N$ and $\u$ are large, pointing to the importance
of multiple mutations in the dynamics.  Clonal interference analysis, by
contrast, assumes that all mutations occur and fix in the wild-type
population, which implies $q$ between $1$ and $2$. This underlies the
calculation of the fixation rate in \eq{clonalfixationrate}, which
assumes that mutations in any individual can contribute to adaptation,
when in fact for $q > 1$ it is only the mutations in the relatively rare
individuals that already have other beneficial mutations which
contribute. Thus we expect that neglecting the importance of multiple
mutations should lead clonal interference analysis to overestimate the
rate of adaptation $v$, for all $\beta$. This is indeed what we find for
$\beta = 0.5$, but we see from Fig. 5 that clonal interference analysis
accurately predicts $v$ for $\beta = 1$, and actually underestimates the
rate of adaptation for $\beta = 10$.

The reason for this discrepancy is apparent from Fig. 3, which shows that
for $\beta = 1$ and $\beta = 10$, clonal interference tends to
underestimate $\rms$.  This problem gets worse as $N$ and $\u$ increase.
The reason for this underestimate is that clonal interference assumes
that the largest mutation that occurs before any other fixes goes to
fixation by itself. This strongly suppresses the fixation of mutations
which have a substantially smaller fitness effect, and leads to a
prediction of a very small $\rms$. But in fact, mutations with a variety
of smaller effects will sometimes happen to occur in individuals that
have this larger-effect mutation, and these will sweep to fixation
together.  This broadens the distributions of mutations that fix, and
hence increases the actual $\rms$, as is apparent in Fig. 3 and in the
distributions shown in Figs. 1b and 1c. This means that clonal
interference assumes that only mutations within a much narrower range of
fitness effects contribute to adaptation than is actually the case, which
should lead to underestimates of the rate of adaptation.  In other words,
although only mutations that happen in very fit individuals can
contribute (which slows adaptation), many mutations of various effects
occur in these fit individuals and can all fix together (speeding
adaptation). This underestimate of $v$ is more severe for larger $\beta$,
because larger $\beta$ corresponds to a larger underestimate of $\rms$.
We see from Fig. 5 that for $\beta = 1$, this underestimate of $v$
roughly cancels the overestimate of $v$ caused by the assumption that
mutations in any individual can contribute to adaptation. For $\beta =
0.5$, the underestimate of $\rms$ is less severe, so it only partially
cancels the overestimate, and in sum clonal interference overestimates
the rate of adaptation.  For $\beta = 10$, the reverse is true.

Together, these results suggest that clonal interference analysis is the
right framework for estimating $\tilde s$ whenever $\beta \leq 1$. For
$\beta = 0.5$, it also gets the distribution of mutations that fix
roughly correct, while for $\beta = 1$ it underestimates $\rms$, and
hence misunderstands the distribution of mutations that actually fix. For
$\beta = 10$, it does not accurately predict either $\tilde s$ or the
shape of the distribution of mutations that fix.  Although it gives
accurate estimates of $v$ for $\beta = 1$, it is apparent that for all
$\beta$ clonal interference remains an incomplete picture of the
dynamics; multiple mutations are also important in a variety of ways.

An alternative framework, which focuses primarily on multiple-mutation
effects, was proposed by \citet{desaifisher07}.  These and other authors
studied a model where all beneficial mutations have the \emph{same}
fitness advantage $s$ \cite{rouzine03,
rouzinewilke07,kesslerlevine98,desaifisher07}. They calculated the rate
at which these mutations accumulate, $v(s)$, as a function of population
size and mutation rate.  \citet{desaifisher07} then argued that in a more
general situation where beneficial mutations have a range of fitness
effects, under many conditions $\rms$ should be small compared to $\tilde
s$, so that mutations within a narrow range of fitness effects dominate
the evolution.  Thus the single-$s$ model describes the full dynamics,
provided that one chooses that single $s$ to be $\tilde s$, and chooses
the beneficial mutation rate to these mutations to be the total mutation
rate towards all mutations within roughly $\rms$ of $\tilde s$.  This
should be true as long as $\rms$ is relatively narrow --- at most of
order $\tilde s$.  Our simulations show that this is indeed the case
(Fig. 3), as do recent experimental studies in \emph{S. cerevisiae} and
\emph{E. coli} \cite{hegreness06, desaimurray07}.

Of course, the value of $\tilde s$ and the width of the range of
mutations which contribute to the evolution depend on $\rho(s)$, $N$, and
the overall mutation rate.  A full understanding of this depends in a
subtle way on both clonal interference and multiple-mutation effects, but
\citet{desaifisher07} proposed a simple approximation.  They first
calculate $v(s)$ for each possible value of $s$, assuming that only
mutations of this size are possible.  To do this, one must specify an
appropriate mutation rate to mutations of this size, which we refer to as
$\tilde \u$. \citet{desaifisher07} made the ad-hoc assumption that this
mutation rate should be the total mutation rate to mutations of order $s$
(i.e. within roughly a factor of $2$ of $s$).  They then calculated
$v(s)$. This $v(s)$ expresses the contribution of mutations of effect $s$
to the overall evolution, and thus should equal $R(s)$, up to
normalization. From this $R(s)$, they calculate $\tilde s$. They find
\eon \tilde s = \sig \left[ \frac{\ell}{\beta - 1} \right]^{1/\beta},
\label{stildemult} \eoff where $\ell$ is related to the overall mutation
rate by \eon \ell = - \ln \left[ \frac{\u}{\sig \Gamma(1+ 1/\beta)}
\right]. \eoff This expression for $\tilde s$ is only valid for $\beta
> 1$; for distributions of mutational effects that fall off exponentially
or slower the behavior is more complicated and the analysis breaks down.

Both this approximation and the original clonal interference analysis
predict that $\tilde s$ should increase with $N$, because increasing the
population size increases the probability of clonal interference but does
not dramatically change the relative importance of large effect mutations
relative to multiple smaller-effect ones.  On the other hand, for large
$\beta$, multiple-mutation effects should cause a qualitative shift in
the relationship between $\tilde s$ and $\u$. Clonal interference
analysis alone predicts that $\tilde s$ increases with $\u$, because
higher mutation rates make clonal interference more common. However,
higher mutation rates also increase the importance of multiple
small-effect mutations relative to large-effect ones. From
\eq{stildemult} we see that for $\beta > 1$ this should mean that $\tilde
s$ actually \emph{decreases} with $\u$ for large $\u$. Although the
effect is small, we do observe this decrease in our simulations (Fig.
2a). For small $\beta$, on the other hand, we expect that multiple
mutation effects typically do not impede the fixation of large-effect
mutations. Thus clonal interference alone gives a qualitatively accurate
picture of how $\tilde s$ depends on $N$ and $\u$, as observed.

Given $\tilde s$, and assuming that the appropriate mutation rate is that
towards mutations of this order, the dynamics of adaptation are similar
to that of the corresponding single-$s$ model.  Using \eq{stildemult} and
their analysis of the single-$s$ model, \citet{desaifisher07} calculated
how the rate of adaptation $v$ and the lead $q$ should depend on
$N$ and $\u$ and the shape of the distribution of mutational effects.
They found \eon \label{qmult} q \approx \frac{2 \ln \left[ N \sig \right]
( \beta - 1 )}{\ell \beta}, \eoff  and \eon \label{vmult} v \approx 2
C_\beta \sig^2 \frac{\ln \left[ N \sig \right]}{\ell^{2 - 2/\beta}},
\qquad \textrm{where} \quad C_\beta \equiv \frac{(\beta -1 )^{2 -
2/\beta}}{\beta^2}.\eoff As with \eq{stildemult}, these expressions are
only valid for $\beta > 1$.

We compare these theoretical predictions to our simulation results in
Figs. 2, 3 and 5.  We see that for large $N$ and $\u$ they give
qualitatively the correct behavior for $\tilde s$, the lead $q$, and the
rate of adaptation $v$, though they do systematically overestimate $q$
and $v$ (see below).  In this regime, these results are more accurate
than those given by clonal interference analysis.  Two qualitative
features are particularly important:  that $q$ can become larger than
$2$, and that $\tilde s$ actually \emph{decreases} as $\u$ increases when
$\beta > 1$.  For small $N$ and $\u$, corresponding to $q \leq 2$, the
multiple-mutations results are less accurate, as expected because the
above results are valid only when $q \gtrsim 2$ \cite{desaifisher07}.

\eq{qmult} and \eq{vmult} rely on the ad-hoc assumption that the
appropriate mutation rate to the mutations around $\tilde s$ that
dominate the dynamics is the total mutation rate to mutations of order
$\tilde s$ (i.e. within roughly a factor of $2$ of $\tilde s$).  As we
have found here, $\rms$ is often much smaller than $\tilde s$, so the
range of mutational effects that contributes substantially to the
evolution is much smaller than assumed.  This means that the intuitive
picture behind the multiple-mutations model of the dynamics is correct:
there is indeed a narrow range of mutations which contribute to the
evolution, and their accumulation is well-described by a corresponding
single-$s$ model of the dynamics. However, the estimate of the
appropriate mutation rate is too large, since a narrower range
contributes than was assumed. This means that the predictions of $q$ and
$v$ in \eq{qmult} and \eq{vmult} should be overestimates, as we observe.
To correct these overestimates, we need to understand what determines
$\rms$.  Unfortunately neither clonal interference analysis nor the
multiple-mutations approach does this well; it remains an important topic
for future analytical work.

While the clonal interference and the multiple-mutations approximations
together help us to form a more complete understanding of the dynamics,
both leave much to be desired.  Clonal interference processes appear to
be the main determinant of $\tilde s$ when $\beta \leq 1$, as
demonstrated by Fig. 2.  But even for these small $\beta$, we see that
$q$ is often larger than $2$, and hence clonal interference analysis
gives the wrong picture for the dynamics.  Further, it misses the
possibility of smaller-effect mutations fixing together with those of
effect roughly $\tilde s$, and hence drastically underestimates $\rms$
even for $\beta = 1$.  On the other hand, while the multiple-mutations
analysis provides the right picture of the dynamics for $\beta > 1$, and
accurately predicts $\tilde s$, $q$ and $v$, it provides no way to
estimate $\rms$, nor any specific predictions when $\beta \leq 1$.  The
simulation approach we have taken in this paper sheds some light on where
and why these two different approaches work, and has highlighted the
regimes where neither provides a satisfactory picture.  A more detailed
understanding will require analysis of a general model which explicitly
incorporates both clonal interference and multiple mutations, to produce
a theory which has the correct picture of the dynamics in these difficult
regimes.

An interesting result of our simulations is that the general shape of the
distribution of the mutations that contribute to adaptation, $R(s)$, is
rather universal. $R(s)$ is always a relatively narrow distribution with
a clear peak at some $\tilde s$.  The distribution of mutations that fix,
$\rho_f(s)$, has a similar shape with a sharp peak near $\tilde s$.  This
means that if we do a single experiment at a given population size and
mutation rate, the observed $\rho_f(s)$ will not provide much information
about the underlying $\rho(s)$.  This lack of sensitivity of experimental
adaptation to the distribution of mutational effects has been noted in a
related context by \citet{hegreness06}. However, our simulations also
show that the scaling of various aspects of $\rho_f(s)$ (such as $\tilde
s$) with population size and mutation rate \emph{does} depend strongly on
$\rho(s)$.  Most important is the shape of the tail of $\rho(s)$; as we
have seen, the way in which $\tilde s$ depends on $N$ and $\u$ is
strongly dependent on $\beta$.  Thus careful experiments which are
carried out at a range of population sizes or mutation rates may make it
possible to infer important aspects of $\rho(s)$ from measurements of
$\rho_f(s)$ or the rate of adaptation $v$.

As our simulations make clear, the actual values of $\beta$ applicable to
natural asexual populations are of central importance to the dynamics by
which these populations adapt.  The values of $\beta$ found in natural
populations may also tell us something about the evolutionary history of
these populations.  Orr and Gillespie have argued that if an individual
is at a random high-fitness genotype, the distribution of more-fit
genotypes is exponential, so we should expect $\beta = 1$ \cite{orr02,
orr03, gillespie83, gillespie84, gillespie91}.  However, as a population
adapts it is natural to expect $\rho(s)$ to change. For example, a
population may face a static challenge, and gradually deplete the
available beneficial mutations as it adapts.  If the population were
small enough that $\rho_f(s) = 2 s \rho(s)$, the distribution $\rho(s)$
should converge to an exponential as this adaptation progresses.  But if
clonal interference and multiple-mutation effects cause large-effect
mutations to be depleted much faster than small-effect ones, we expect
that $\beta$ should increase as the available mutations become depleted.
Thus large observed values of $\beta$ may be indicative of this type of
adaptation. Of course, this increase in $\beta$ could be avoided if the
accumulation of beneficial mutations tends to open up new possibilities
for further adaptation.  In other words, the structure of fitness
landscapes has an important role in determining typical values of
$\beta$, and how these change as populations adapt.  Given these values
of $\beta$, our simulations provide a way to understand, in a statistical
sense, which mutations will tend to contribute to adaptation, and the
dynamics by which they will do so.

\section{Acknowledgments}
JLN and CF acknowledge support from the University of Colorado and the
Undergraduate Research Opportunity Program at the University of Colorado.
MMD acknowledges support from Center grant P50GM071508 from the National
Institute of General Medical Science to the Lewis-Sigler Institute.

\clearpage
\newpage

\bibliographystyle{genetics}
\bibliography{./mutationmc}

\begin{thebibliography}{26}

\bibitem[{\sc de~Visser {\em et~al.\/}}(1999){\sc de~Visser, Zeyl, Gerrish,
  Blanchard, {\rm and} Lenski}]{devisser99}
{\sc de~Visser, J., C.~W. Zeyl, P.~J. Gerrish, J.~L. Blanchard, {\rm and} R.~E.
  Lenski}, 1999 Diminishing returns from mutation supply rate in asexual
  populations. Science {\bf 283}: 404--406.

\bibitem[{\sc de~Visser {\rm and} Rozen}(2005)]{devisserrozen05}
{\sc de~Visser, J. A. G.~M. {\rm and} D.~E. Rozen}, 2005 Limits to adaptation
  in asexual populations. Journal of Evolutionary Biology {\bf 18}: 779--788.

\bibitem[{\sc DePristo {\em et~al.\/}}(2005){\sc DePristo, Weinreich, {\rm and}
  Hartl}]{depristo05}
{\sc DePristo, M.~A., D.~M. Weinreich, {\rm and} D.~Hartl}, 2005 Missense
  meanderings in sequence space: A biophysical view of protein evolution.
  Nature Reviews Genetics {\bf 6}: 678--687.

\bibitem[{\sc Desai {\rm and} Fisher}(2007)]{desaifisher07}
{\sc Desai, M.~M. {\rm and} D.~S. Fisher}, 2007 Beneficial mutation-selection
  balance and the effect of linkage on positive selection. Genetics {\bf 176}:
  1759--1798.

\bibitem[{\sc Desai {\em et~al.\/}}(2007){\sc Desai, Fisher, {\rm and}
  Murray}]{desaimurray07}
{\sc Desai, M.~M., D.~S. Fisher, {\rm and} A.~W. Murray}, 2007 The speed of
  evolution and maintenance of variation in asexual populations. Current
  Biology {\bf 17}: 385--394.

\bibitem[{\sc Gerrish}(2001)]{gerrish01}
{\sc Gerrish, P.}, 2001 The rhythm of microbial adaptation. Nature {\bf 413}:
  299--302.

\bibitem[{\sc Gerrish {\rm and} Lenski}(1998)]{gerrishlenski98}
{\sc Gerrish, P. {\rm and} R.~Lenski}, 1998 The fate of competing beneficial
  mutations in an asexual population. Genetica pp. 127--144.

\bibitem[{\sc Gillespie}(1983)]{gillespie83}
{\sc Gillespie, J.~H.}, 1983 A simple stochastic gene substitution model.
  Theoretical Population Biology {\bf 23}: 202--215.

\bibitem[{\sc Gillespie}(1984)]{gillespie84}
{\sc Gillespie, J.~H.}, 1984 Molecular evolution over the mutational landscape.
  Evolution {\bf 38}: 1116--1129.

\bibitem[{\sc Gillespie}(1991)]{gillespie91}
{\sc Gillespie, J.~H.}, 1991 {\em The Causes of Molecular Evolution\/}. Oxford
  Univeristy Press, Oxford.

\bibitem[{\sc Haldane}(1927)]{haldane27}
{\sc Haldane, J. B.~S.}, 1927 The mathematical theory of natural and artificial
  selection, part v: Selection and mutation. Proceedings of the Cambridge
  Philosophical Society {\bf 23}: 838--844.

\bibitem[{\sc Hegreness {\em et~al.\/}}(2006){\sc Hegreness, Shoresh, Hartl,
  {\rm and} Kishony}]{hegreness06}
{\sc Hegreness, M., N.~Shoresh, D.~Hartl, {\rm and} R.~Kishony}, 2006 An
  equivalence principle for the incorporation of favorable mutations in asexual
  populations. Science {\bf 311}: 1615--1617.

\bibitem[{\sc Hill {\rm and} Robertson}(1966)]{hillrobertson66}
{\sc Hill, W.~G. {\rm and} A.~Robertson}, 1966 Effect of linkage on limits to
  artificial selection. Genetical Research {\bf 8}: 269--294.

\bibitem[{\sc Imhof {\rm and} Schlotterer}(2001)]{imhof01}
{\sc Imhof, M. {\rm and} C.~Schlotterer}, 2001 Fitness effects of advantageous
  mutations in evolving escherichia coli populations. PNAS {\bf 98}:
  1113--1117.

\bibitem[{\sc Kassen {\rm and} Bataillon}(2006)]{kassen06}
{\sc Kassen, R. {\rm and} T.~Bataillon}, 2006 Distribution of fitness effects
  among beneficial mutations before selection in experimental populations of
  bacteria. Nature Genetics {\bf 38}: 484--488.

\bibitem[{\sc Kim {\rm and} Orr}(2005)]{kimorr05}
{\sc Kim, Y. {\rm and} H.~A. Orr}, 2005 Adaptation in sexuals vs. asexuals:
  clonal interference and the fisher-muller model. Genetics {\bf 171}:
  1377--1386.

\bibitem[{\sc Lunzer {\em et~al.\/}}(2005){\sc Lunzer, Miller, Felsheim, {\rm
  and} Dean}]{lunzer05}
{\sc Lunzer, M., S.~P. Miller, R.~Felsheim, {\rm and} A.~M. Dean}, 2005 The
  biochemical architecture of an ancient adaptive landscape. Science {\bf 310}:
  499--501.

\bibitem[{\sc Orr}(2002)]{orr02}
{\sc Orr, H.~A.}, 2002 The population genetics of adaptation: The adaptation of
  dna sequences. Evolution {\bf 56}: 1317--1330.

\bibitem[{\sc Orr}(2003)]{orr03}
{\sc Orr, H.~A.}, 2003 The distribution of fitness effects among beneficial
  mutations. Genetics {\bf 163}: 1519--1526.

\bibitem[{\sc Ridgway {\em et~al.\/}}(1998){\sc Ridgway, Levine, {\rm and}
  Kessler}]{kesslerlevine98}
{\sc Ridgway, D., H.~Levine, {\rm and} D.~Kessler}, 1998 Evolution on a smooth
  landscape: the role of bias. Journal of Statistical Physics {\bf 90}: 191.

\bibitem[{\sc Rokyta {\em et~al.\/}}(2005){\sc Rokyta, Joyce, Caudle, {\rm and}
  Wichman}]{rokyta05}
{\sc Rokyta, D.~R., P.~Joyce, S.~B. Caudle, {\rm and} H.~A. Wichman}, 2005 An
  empirical test of the mutational landscape model of adaptation using a
  single-stranded dna virus. Nature Genetics {\bf 37}: 441--444.

\bibitem[{\sc Rouzine {\em et~al.\/}}(2008){\sc Rouzine, Brunet, {\rm and}
  Wilke}]{rouzinewilke07}
{\sc Rouzine, I., E.~Brunet, {\rm and} C.~O. Wilke}, 2008 The traveling-wave
  approach to asexual evolution: Muller's ratchet and speed of adaptation.
  Theoretical Population Biology {\bf in press}.

\bibitem[{\sc Rouzine {\em et~al.\/}}(2003){\sc Rouzine, Wakeley, {\rm and}
  Coffin}]{rouzine03}
{\sc Rouzine, I., J.~Wakeley, {\rm and} J.~M. Coffin}, 2003 The solitary wave
  of asexual evolution. PNAS {\bf 100}: 587--592.

\bibitem[{\sc Rozen {\em et~al.\/}}(2002){\sc Rozen, de~Visser, {\rm and}
  Gerrish}]{rozen02}
{\sc Rozen, D.~E., J.~de~Visser, {\rm and} P.~J. Gerrish}, 2002 Fitness effects
  of fixed beneficial mutations in microbial populations. Current Biology {\bf
  12}: 1040--1045.

\bibitem[{\sc Sanjuan {\em et~al.\/}}(2004){\sc Sanjuan, Moya, {\rm and}
  Elena}]{sanjuan04}
{\sc Sanjuan, R., A.~Moya, {\rm and} S.~F. Elena}, 2004 The distribution of
  fitness effects caused by single-nucleotide substitutions in an rna virus.
  Proceedings of the National Academy of Sciences of the United States of
  America {\bf 101}: 8396--8401.

\bibitem[{\sc Wilke}(2004)]{wilke04}
{\sc Wilke, C.~O.}, 2004 The speed of adapation in large asexual populations.
  Genetics {\bf 167}: 2045--2053.

\end{thebibliography}

\newpage

\baselineskip24pt

\linespread{1.0}

\begin{figure}
  \includegraphics[width=\textwidth]{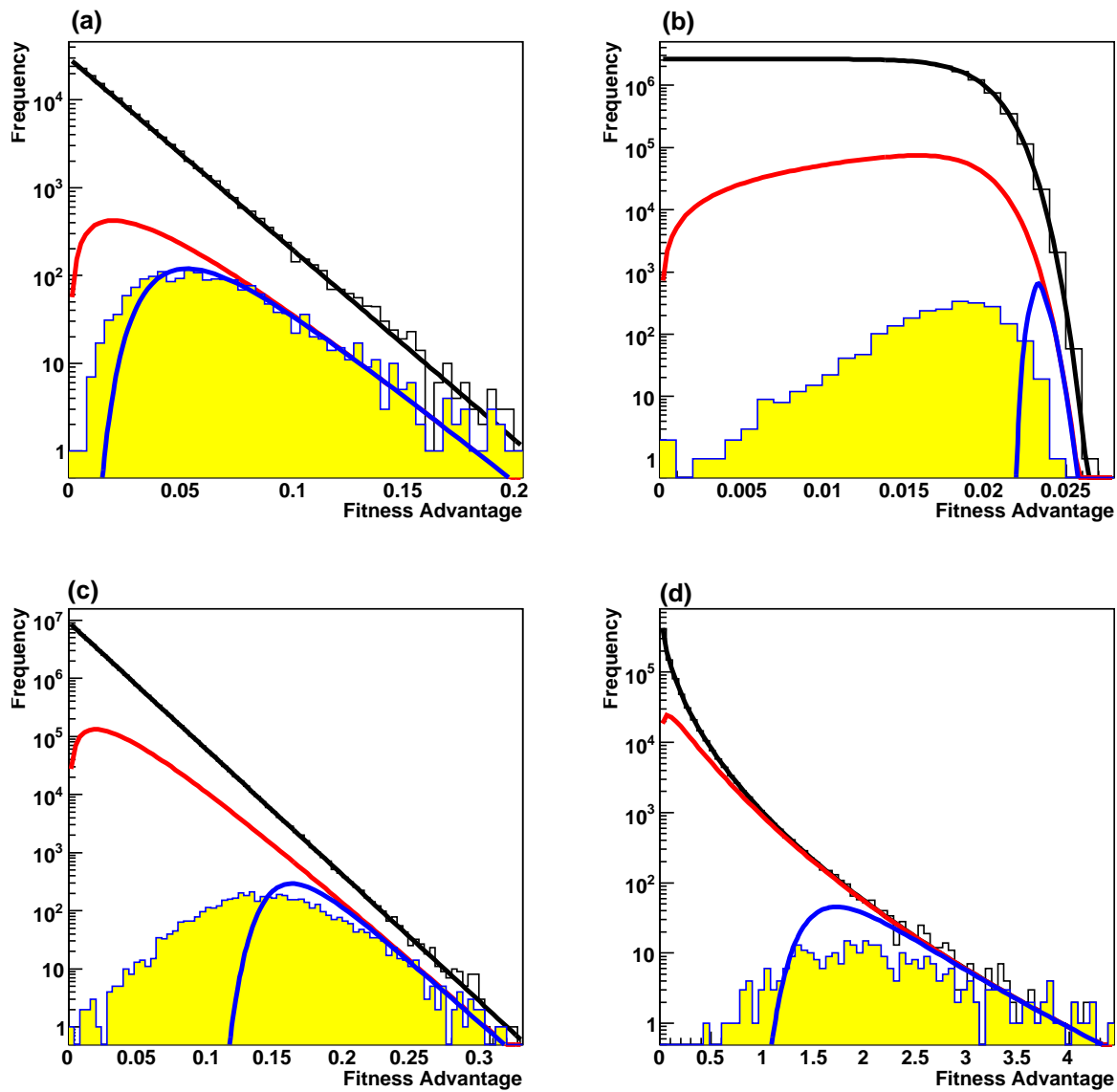}
  \caption{\label{1a}
Examples of the distribution of possible mutations $\rho(s)$ (solid black
line), the distribution of contending mutations $\rho_c(s)$ (solid red
line), and the distribution of fixed mutations $\rho_f(s)$ from our
simulations (yellow histogram).  Also shown is the distribution of
mutations which occurred in the simulations (transparent histogram).
Additionally the distribution of predicted fixed mutations $\rho_f(s)$
from a clonal interference calculation are shown (solid blue line).  Note
the logarithmic scale.  In all examples, $\sigma = 0.02$. (\textbf{a}) A
small population, where $\rho_f(s) = \rho_c(s)$ except for the
smallest-effect mutations. Here $N = 3 \times 10^4$, $\u = 10^{-5}$, and
$\beta = 1.0$.  (\textbf{b}) A large population with $\beta = 10$. Here
$N=10^{7}$ and $U_{b} = 10^{-5}$. Note that small-effect mutations are
suppressed by clonal interference effects, while large-effect mutations
are suppressed by multiple mutation effects. (\textbf{c})  A large
population with $\beta = 1$.  Here $N=10^{7}$ and $U_{b} = 10^{-5}$. Note
that small-effect mutations are suppressed by clonal interference
effects, but less strongly than clonal interference analysis alone
predicts.  (\textbf{d}) A large population with $\beta = 0.5$. Here $N=1
\times 10^{7}$ and $U_{b} = 10^{-5}$. }
\end{figure}

\begin{figure}
  \includegraphics[width=\textwidth]{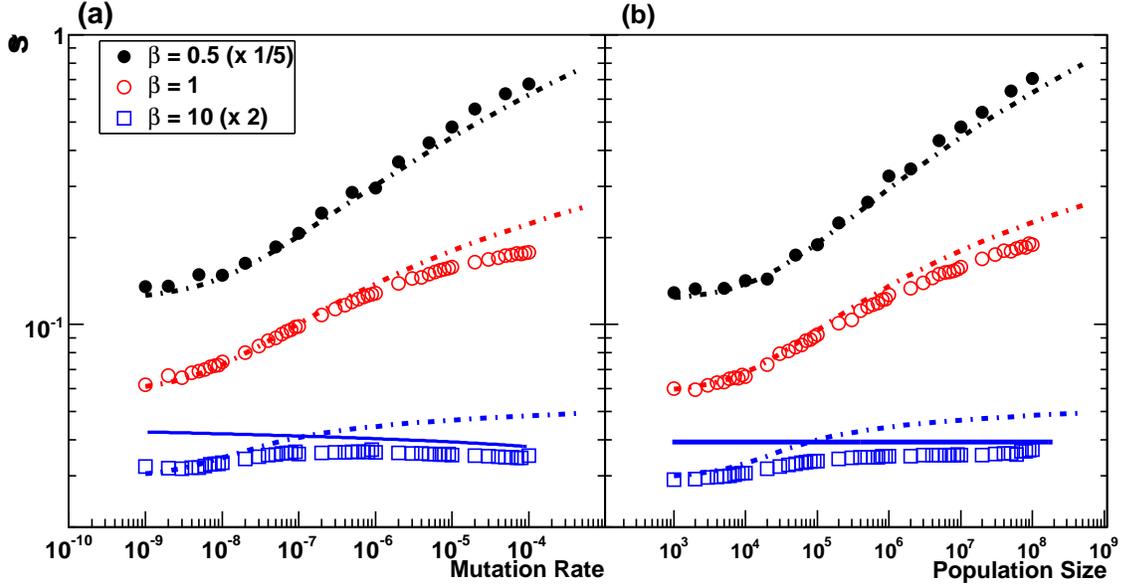}
  \caption{\label{2}
The average scaled fitness effect of mutations that fix, $\tilde s$. In
all cases $\sigma = 0.02$.  (\textbf{a}) Simulation results for $\tilde
s$ for $N = 10^7$ as a function of $\u$ for $\beta = 0.5$ (closed black
circles), $\beta = 1$ (open red circles), and $\beta = 10$ (open blue
squares). Predictions of clonal interference analysis are shown as dotted
lines, and the predictions of the multiple mutation analysis (for $\beta
= 10$) are shown as a solid line. Note that for $\beta = 10$, $\tilde s$
decreases with $\u$ for large $\u$. (\textbf{b}) Simulation results for
$\tilde s$ for $\u = 10^{-5}$ as a function of $N$ for $\beta = 0.5$
(closed black circles), $\beta = 1$ (open red circles), and $\beta = 10$
(open blue squares). Predictions of clonal interference analysis are
shown as dotted lines, and the predictions of the multiple mutation
analysis (for $\beta = 10$) are shown as a solid line. }
\end{figure}

\begin{figure}
  \includegraphics[width=\textwidth]{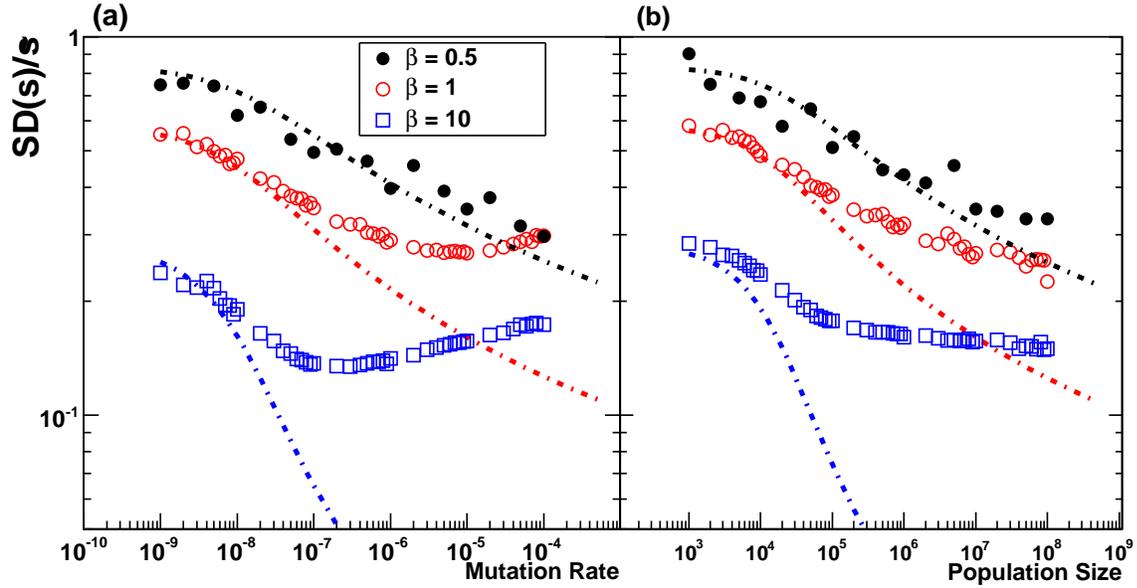}
  \caption{\label{3}
The scaled width of the weighted distribution of mutations that fix,
$\rms/ \tilde s$.  In all cases $\sigma = 0.02$.  (\textbf{a}) Simulation
results for $\rms/ \tilde s$ for $N = 10^7$ as a function of $\u$ for
$\beta = 0.5$, $\beta = 1$, and $\beta = 10$. Predictions of clonal
interference analysis are shown as dotted lines. (\textbf{b}) Simulation
results for $\tilde s$ for $\u = 10^{-5}$ as a function of $N$ for $\beta
= 0.5$, $\beta = 1$, and $\beta = 10$. Predictions of clonal interference
analysis are shown as dotted lines. }
\end{figure}

\newpage
\clearpage

\begin{figure}
  \includegraphics[width=\textwidth]{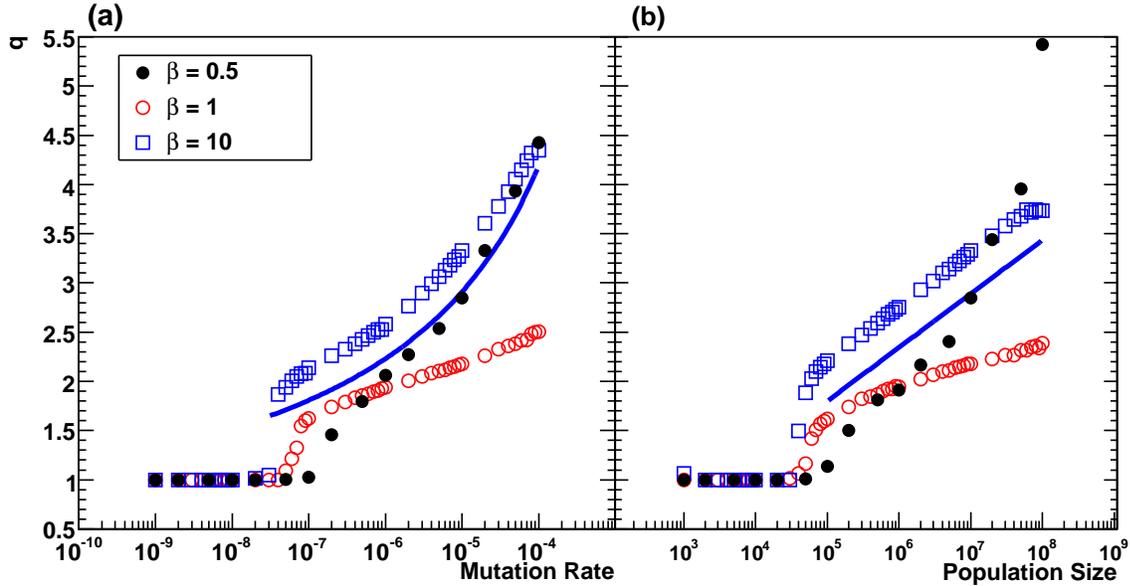}
  \caption{\label{4}
The effective lead $q$.  (\textbf{a}) Simulation results for $q$ as a
function of $\u$ for $\beta = 0.5$, $\beta = 1$, and $\beta = 10$. Other
parameters are as in Fig. 3a.  Predictions of the multiple-mutations
analysis (for $\beta = 10$) are shown as a solid blue line.  Note that
clonal interference predicts $q \approx 1$, independent of the
parameters. (\textbf{b}) Simulation results for $q$ as a function of $N$
for $\beta = 0.5$, $\beta = 1$, and $\beta = 10$. Other parameters are as
in Fig. 3b.  Predictions of the multiple-mutations analysis (for $\beta =
10$) are shown as a solid blue line.  As before, clonal interference
predicts $q \approx 1$, independent of the parameters. }
\end{figure}

\begin{figure}
  \includegraphics[width=\textwidth]{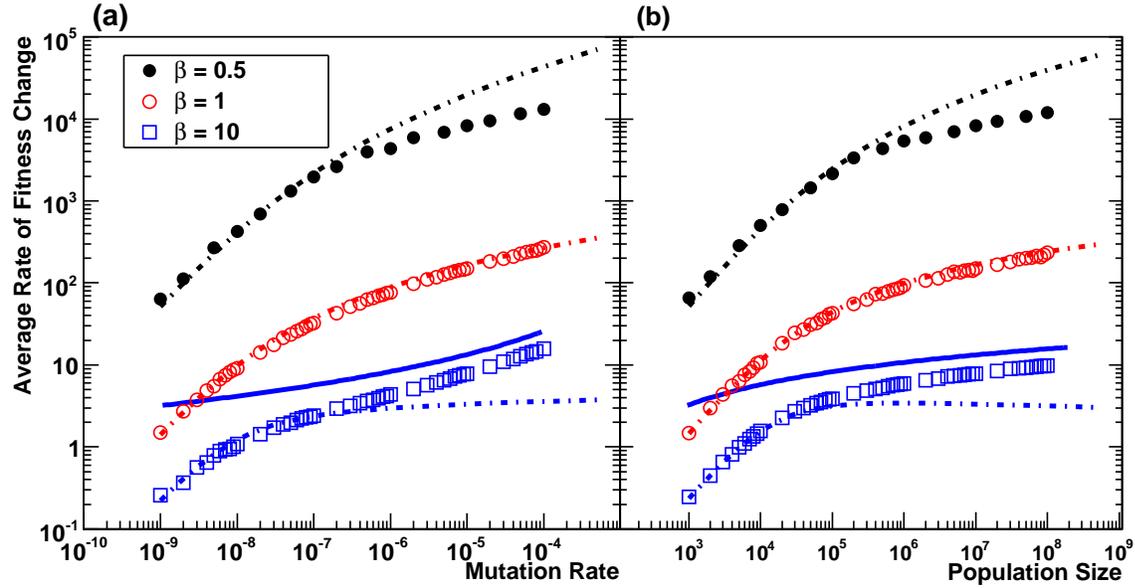}
  \caption{\label{5}
The average rate of adaptation, $v$.  (\textbf{a}) Simulation results for
$v$ as a function of $\u$ for $\beta = 0.5$, $\beta = 1$, and $\beta =
10$. Other parameters are as in Fig. 3a.  Predictions of clonal
interference analysis are shown as dotted lines, and the predictions of
the multiple-mutations analysis (for $\beta = 10$) are shown as a solid
blue line. (\textbf{b}) Simulation results for $v$ as a function of $N$
for $\beta = 0.5$, $\beta = 1$, and $\beta = 10$. Other parameters are as
in Fig. 3b.  Predictions of clonal interference analysis are shown as
dotted lines, and the predictions of the multiple-mutations analysis (for
$\beta = 10$) are shown as a solid blue line. }
\end{figure}

\end{document}